\documentclass{paper}
\usepackage{epsfig, graphicx}
\usepackage[english]{babel}

\begin{document}
\title{A Wideband Spectrometer for NMR Studies}
\author{Aleksey M. Tikhonov$^a$\/\thanks{tikhonov@kapitza.ras.ru}, Aleksander Yu. Semanin$^{a,b}$,\\ and  Georgiy D. Sokolov$^{a,b}$}
\maketitle
\leftline{\it $^a$ Kapitza Institute for Physical Problems, Russian Academy of Sciences,}
\leftline{\it ul. Kosygina 2, Moscow, 119334, Russia}
\leftline{\it $^b$ Moscow Institute of Physics and Technology,} 
\leftline{\it Institutskii per. 9, Dolgoprudnyi, Moscow oblast, 141700, Russia}
\vspace{0.1in}

\rightline{\today}

\abstract{The design of a wideband decimeter-wave (200 - 900 MHz) spectrometer with a magnetic induction of up to $\sim 10$\,T is described. This spectrometer is intended for studying electronic - nuclear oscillations in
antiferromagnets at low temperatures (4.2 - 1.3 K). Critical field $H_c = 2.5 \pm 0.3$ T of a reorientation transition
in a noncollinear antiferromagnet Mn$_3$Al$_2$Ge$_3$O$_{12}$ at temperature $T \approx 1.3$ K was determined from a $^{55}{\rm Mn}^{2+}$} NMR spectrum.
\vspace{0.25in}

\large

The special feature of resonance properties of antiferromagnets with $^{55}{\rm Mn}^{2+}$ magnetic ions 
($100 \%$ isotopic composition) at temperatures $T \sim  1$ K consists in the correlation of nuclear 
oscillations with oscillations of the electronic system, which leads to a strong frequency - field dependence 
of the NMR spectrum or to the dynamic shift of its frequency \cite{1,2,3}. 
In easy-plane antiferromagnets with a linear (over the field) mode of
the antiferromagnetic resonance (e.g., MnCO$_3$), the dynamic shift of the NMR frequency is observed in
magnetic fields up to $H \sim 0.5$ T \cite{4}. In noncollinear antiferromagnets (e.g., CsMnBr$_3$), electronic and
nuclear oscillations interact in a wider magnetic field range ($\sim 4$ T) \cite{5,6}.

The NMR spectrum of magnetic ions contains information on the structure of the ground state
of an antiferromagnet, phase transitions in it, and its low-frequency spin dynamics \cite{7}. 
The frequency band, in which the NMR spectrum is usually recorded, is 200 - 700 MHz. A high resonance frequency 
($\gamma_n H_n \sim$ 600 - 700 MHz) in magnetically ordered materials with Mn$^{2+}$ ions is caused by an 
enormous average local field at a nucleus ($H_n \sim 60$ T),
which is basically determined by hyperfine interaction of nuclear and ion spins (gyromagnetic ratio $\gamma_n \approx
10.6$ MHz/T for $^{55}$Mn). The NMR signal intensity is mainly determined by the amplification effect
of the radio-frequency (RF) field by the transverse component of the hyperfine field \cite{8}, making it possible to observe NMR signals when the external radio-frequency field is polarized {\bf h} $||$ {\bf H} \cite{9,10}.

To satisfy the wideband measurement requirements, it is convenient to use a continuous NMR circuit with a high-Q tunable cavity \cite{11}. In this work, we describe the design of the decimeter-wave (wavelength
$\lambda$ varies from 30 to 100 cm) spectrometer, which is the upgraded version of the instrument described 
in \cite{5,6}. The analog automatic frequency control (AFC) system of the old spectrometer has been replaced 
with the digital AFC system, and a fundamentally new cavity is used.

A broadbandness of the spectrometer is, first of all, ensured by two types of tunable resonance systems.
The first system is based on the earlier designed modification of the split-ring cavity \cite{12}. 
Its design is sketched in Fig. 1. The cavity housing in Fig. 1a is a $10 \times 30 \times 8$ mm parallelepiped 
($8 \times 20 \times 11$ mm parallelepiped in Fig. 1b), which is made of oxygen-free copper. 
In the split ring, the inductance is an 8-mm-diameter through hole, 
and the capacitance is a narrow slot (the gap is $ \sim 0.1$ mm wide) in the cavity housing.
The slot length $L_0$ specifies the cavity's eigenfrequency $\nu_0$, since $\nu_0 \sim L_0^{-1/2}$.

\begin{figure}
\hspace{0.5in}
\epsfig{file=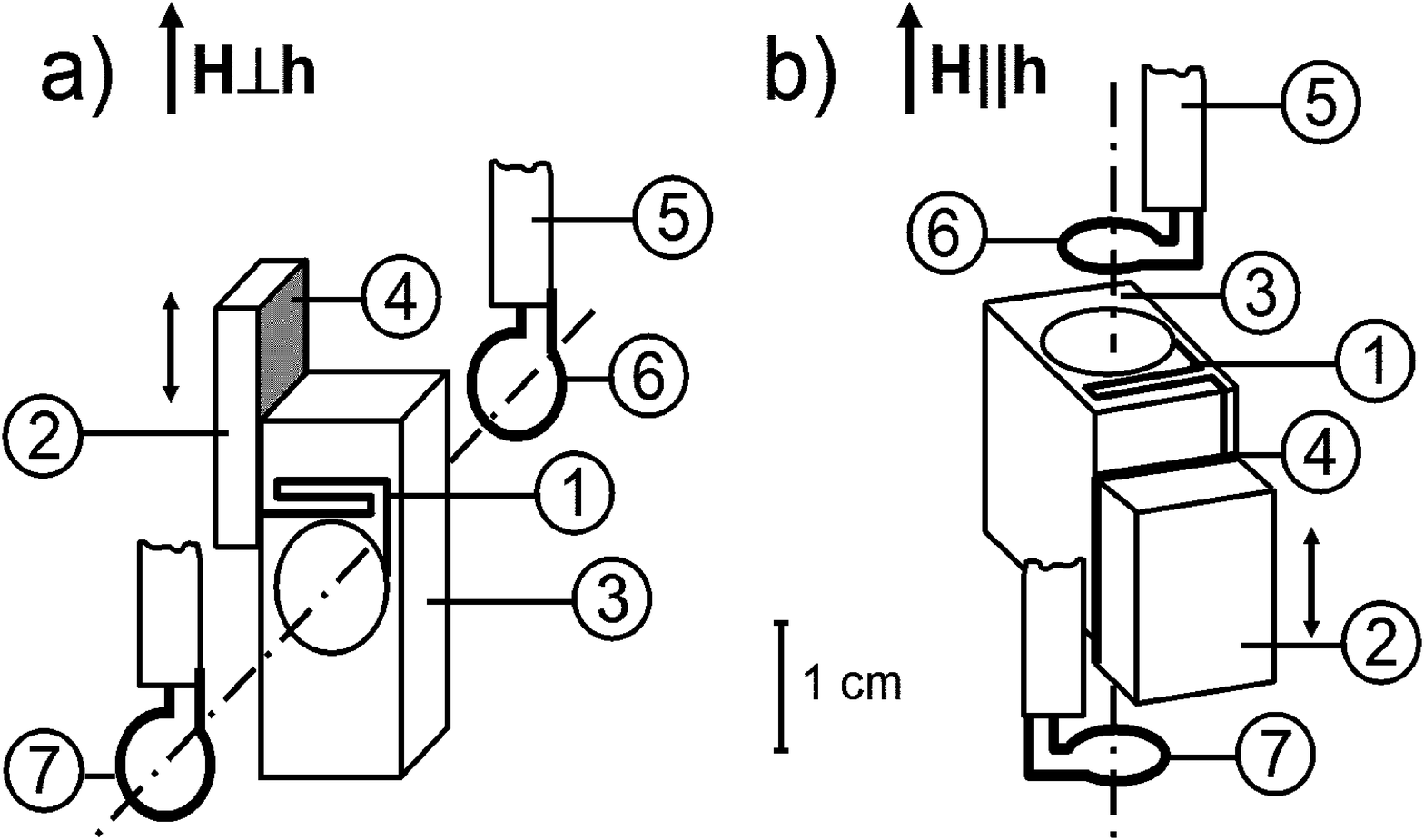, width=0.8
\textwidth}

\vspace{0.1in}
Figure 1. Polarization of the RF field in the split-ring resonance system: (1) narrow slot, (2) copper plate, (3) cavity, (4) polyethylene terephthalate film, (5) coaxial line, and (6, 7) coupling loops.
\end{figure}

The spectrometer frequency is tuned in a $\sim$ 200- to 900-MHz band using three split rings with various
geometries of slot 1, into which mica plates with a 50- to 100-$\mu m$ thickness are 
placed to finely tune the eigenfrequency (see Fig. 2). By shifting copper plate 2, it is possible to change the capacitance between the plate and cavity 3 (isolator 4 is a 5- to 10-$\mu m$-thick
polyethylene terephthalate film), which is used to tune the resonance frequency of the system. 
The RF power is supplied via coaxial line 5. The inductive coupling
with the cavity is affected by means of coupling loops, one of which is transmitting loop 6 
and the other is receiving loop 7. The $\sim$ 5 - mm diameter coupling
loops are placed at a $\sim$ 5 - mm distance from the faces of the cavity (weak coupling). 
The loaded Q factor of the resonance system in the band under investigation
depends on the frequency and varies from 200 to 400 at the liquid helium temperature. 
Depending on the orientation of the cavity axis with respect to the magnetic field $\bf H$ (see Fig. 1), 
it is possible to perform experiments at two polarizations of RF field $\bf h$: 
(a) {\bf h}$\perp${\bf H} and (b) {\bf h}$||${\bf H}.

In contrast to the first system, the second-type resonance system features a significantly higher Q
($\sim$ 3000, see Fig. 3). It is based on a closed cavity, which is a shorted coaxial line 100 mm long 
($\sim \lambda/4$). The coaxial cavity consists of copper cylinder 1 (20-mm inner diameter), 
copper core 2 (8-mm diameter), and bronze membrane 3 with a 0.2-mm thickness. 
Three is a narrow gap (0.2 - 0.5 mm) between the core and membrane. Its value can be changed by exerting 
pressure on the membrane using rod 4. The cavity is coupled by coaxial lines 5 through 2-mm-diameter holes
in the cylinder wall.

\begin{figure}
\hspace{0.5in}
\epsfig{file=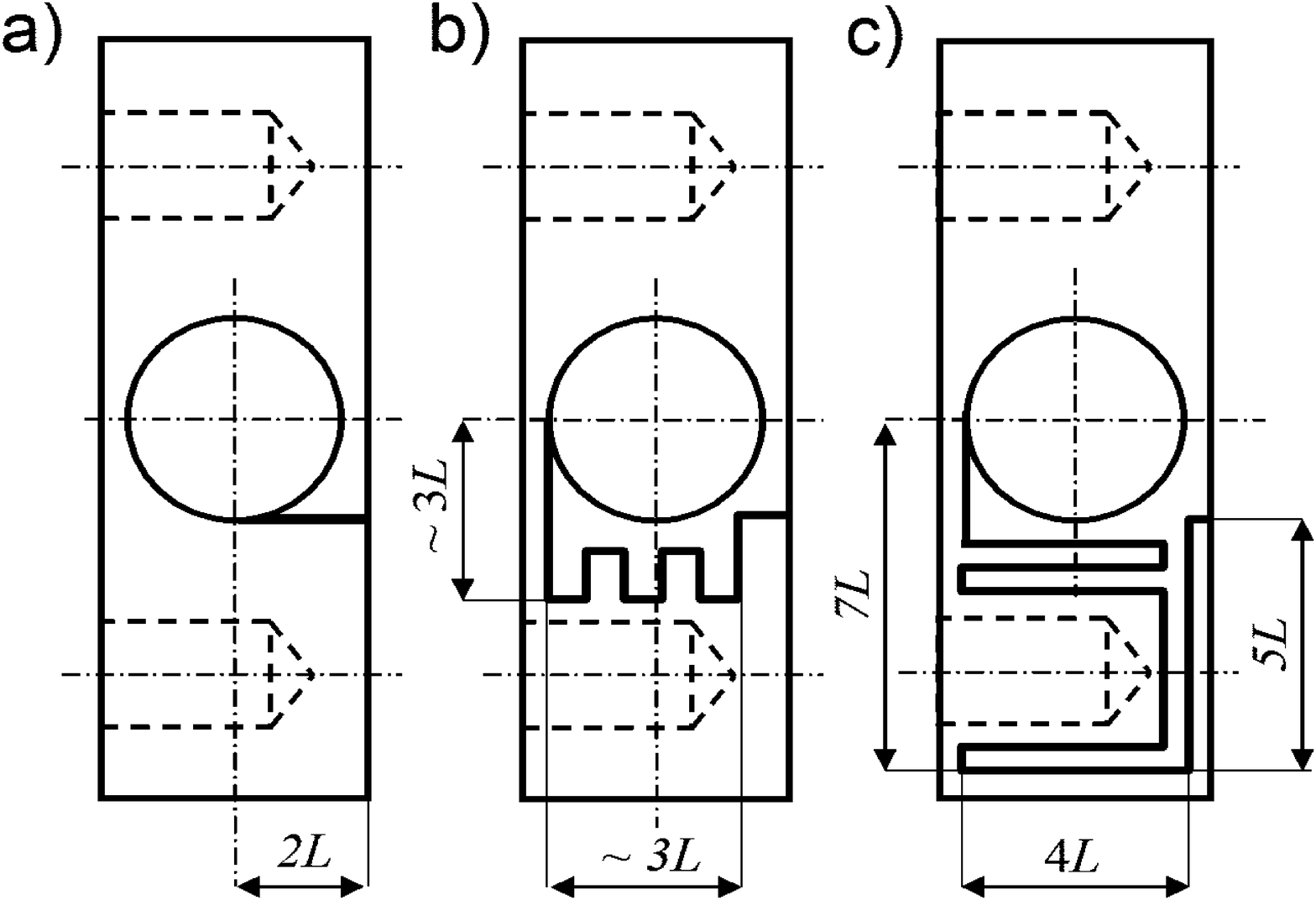, width=0.75\textwidth}

\vspace{0.1in}
Figure 2. Geometric parameters of slots in the split-ring cavities (polarization {\bf h} $\perp$ {\bf H}) with eigenfrequency $\nu_0$: (a) $\sim$ 900 MHz, (b) $\sim$ 450 MHz, and (c) $\sim$ 250 MHz. The through hole diameter is 8 mm, L = 2 mm.
\end{figure}

The lines of the RF magnetic field in the coaxial resonator are the concentric circumferences with a
center lying on the axis of the cavity. The amplitude of this field assumes the minimal value on the membrane
and the maximal value on the lower wall of the cavity. Flat coupling loops 6 are oriented in a radial direction
at right angle to force lines of the RF magnetic field in the upper part of the cavity, where its amplitude is
small, and a sufficiently weak coupling can be arranged.

The external magnetic field is applied along the axis of the cavity. 
Crystal sample 7 is placed at the bottom of the cavity at a maximum of the
RF and stationary magnetic field ({\bf h} $\perp$ {\bf H}), which is created by superconducting solenoid 8 
(with an 80-mm outer diameter and 25-mm inner diameter). The critical current of
the solenoid is $\sim 67$ A at a $\sim 9.7$ T maximal induction.
The intensity of the magnetic field in the experiment is determined from the 
current produced by the bipolar power source (Cryomagnetics-4G-100). The calculated field 
inhomogeneity in the center of the solenoid is $\pm 0.1 \%$ in 1 cm$^3$.
The entire assembly is in a helium bath (glass Dewar flask with a 90-mm inner diameter). 
Its temperature is controlled by a $^4$He equilibrium saturated
vapor pressure regulator with an accuracy of $\pm 0.05$ K or better.

At the liquid helium temperature, the eigenfrequency of the system with the coaxial cavity can vary
from $\sim$ 600 to 625 MHz, and $\nu_0$ of the split-ring system
with the cavity shown in Fig. 2a can be tuned in a $\sim$900- to 500-MHz band.
When the magnetic field increases to maximal values, the frequency stability of the first-type resonance
system is $\sim$ 0.1 MHz, while the second resonance system allows one to fix the frequency in the experiment
with a $\sim$ 10-kHz accuracy. This is likely to be attributed to a high mechanical rigidity of the coaxial cavity.

A block diagram of the spectrometer is shown in Fig. 4. The frequency of the RF signal generator G
(Agilent N9310A) is modulated by a low frequency($f_m$ = 25 kHz) of the auxiliary generator of the 
phase-sensitive voltmeter L1 (SR 830 lock-in amplifier). 
The frequency modulation depth of 0.1 MHz (the maximal frequency deviation resolved by the generator) 
is significantly smaller than the typical NMR line width, which is usually $\>1$ MHz. 
To stabilize the generator frequency on the top of the resonance peak, the AFC
system tuned to zero amplitude of the first modulation harmonic is used. 
The AFC system includes a synchronous detector and digital tracking system
operating in the LabView graphic programming environment (National Instruments). 
When the generator frequency is detuned from the eigenfrequency of the
cavity, signal $U_f$ appears on detector $D$ (planar diode) at the modulation frequency (first harmonic) 
with the phase depending on a mismatch sign. The amplitude of this signal detected by phase-sensitive 
voltmeter L1 is, as a first approximation, proportional to the mismatch value between the generator 
frequency and the eigenfrequency of the resonance system (RS). 
The digital tracking system uses the detected mismatch signal for calculating correction $d\Omega$ to the carrier 
frequency of the RF signal generator as follows:

\begin{figure}
\hspace{0.5in}
\epsfig{file=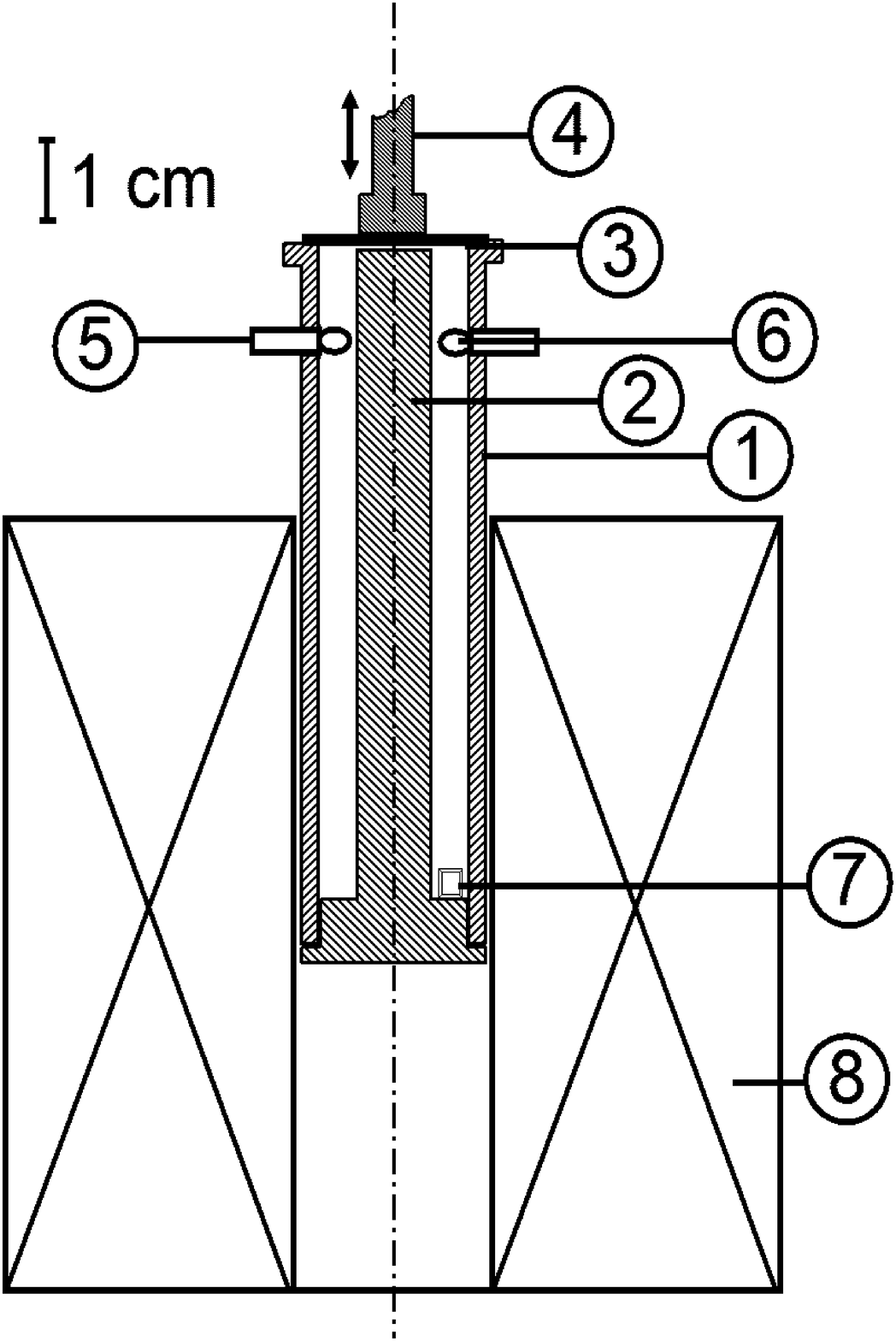, width=0.75\textwidth}

\vspace{0.1in}
Figure 3. Low-temperature part of the NMR spectrometer with a coaxial cavity: (1) cavity walls, (2) core, 
(3) membrane, (4) rod, (5) coaxial line, (6) coupling loop, (7) sample, and (8) solenoid.
\end{figure}

\begin{figure}
\hspace{0.5in}
\epsfig{file=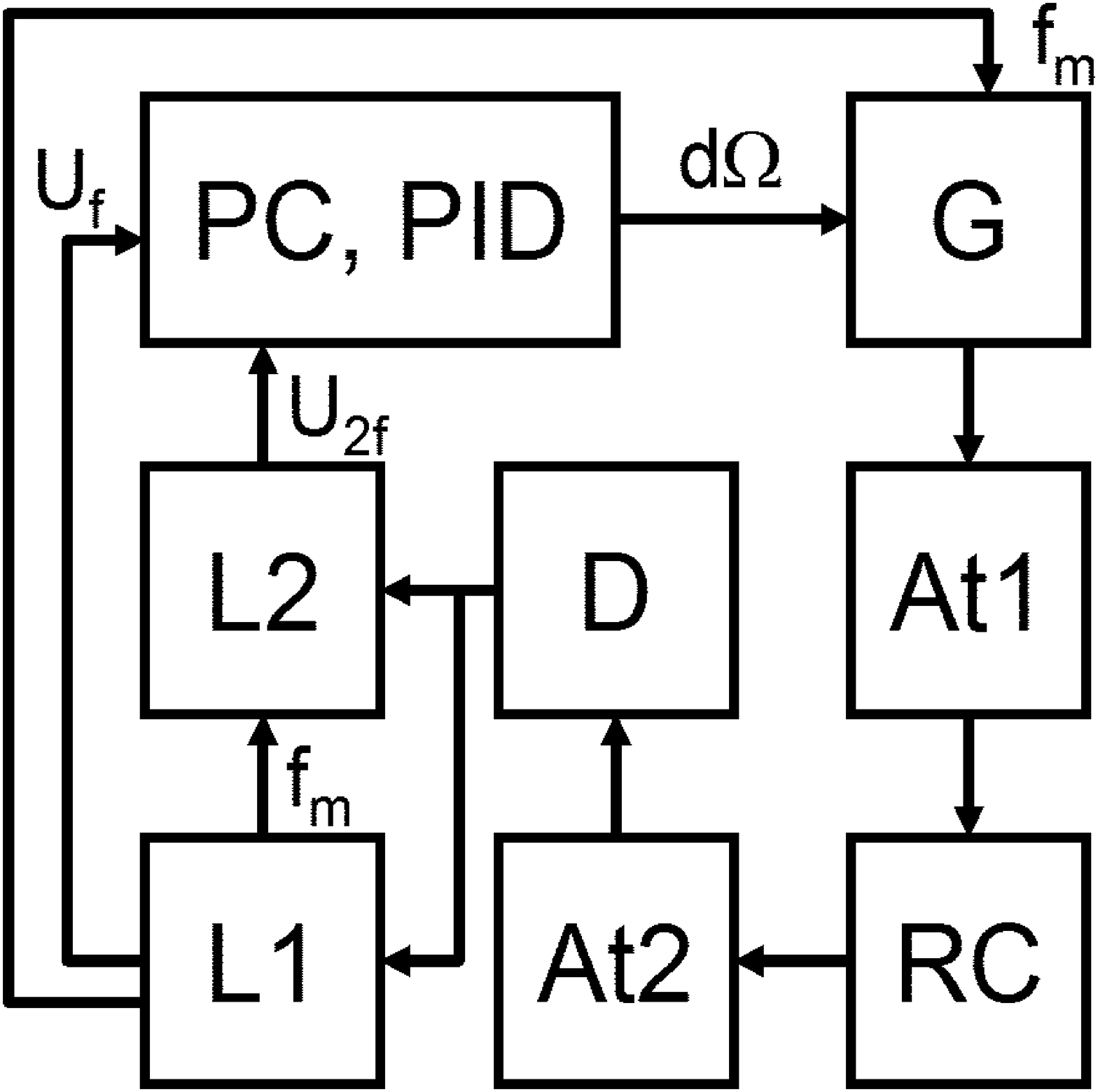, width=0.75\textwidth}

\vspace{0.1in}
Figure 4. Block diagram of the spectrometer. All devices are
integrated into a common control system in the LabView
graphic programming environment (National Instruments):
(RS) resonance system, (G) RF signal generator, (D) planar diode, 
(L1, L2) phase-sensitive voltmeters, (At1, At2)
attenuators, (PC) personal computer, (PID) feedback regulator calculating correction value 
$d\Omega$ to the generator carrier, ($f_m$) modulation signal, ($U_f$) signal of the first harmonic, and
($U_{2f}$) signal of the second harmonic.
\end{figure}

\begin{figure}
\hspace{0.5in}
\epsfig{file=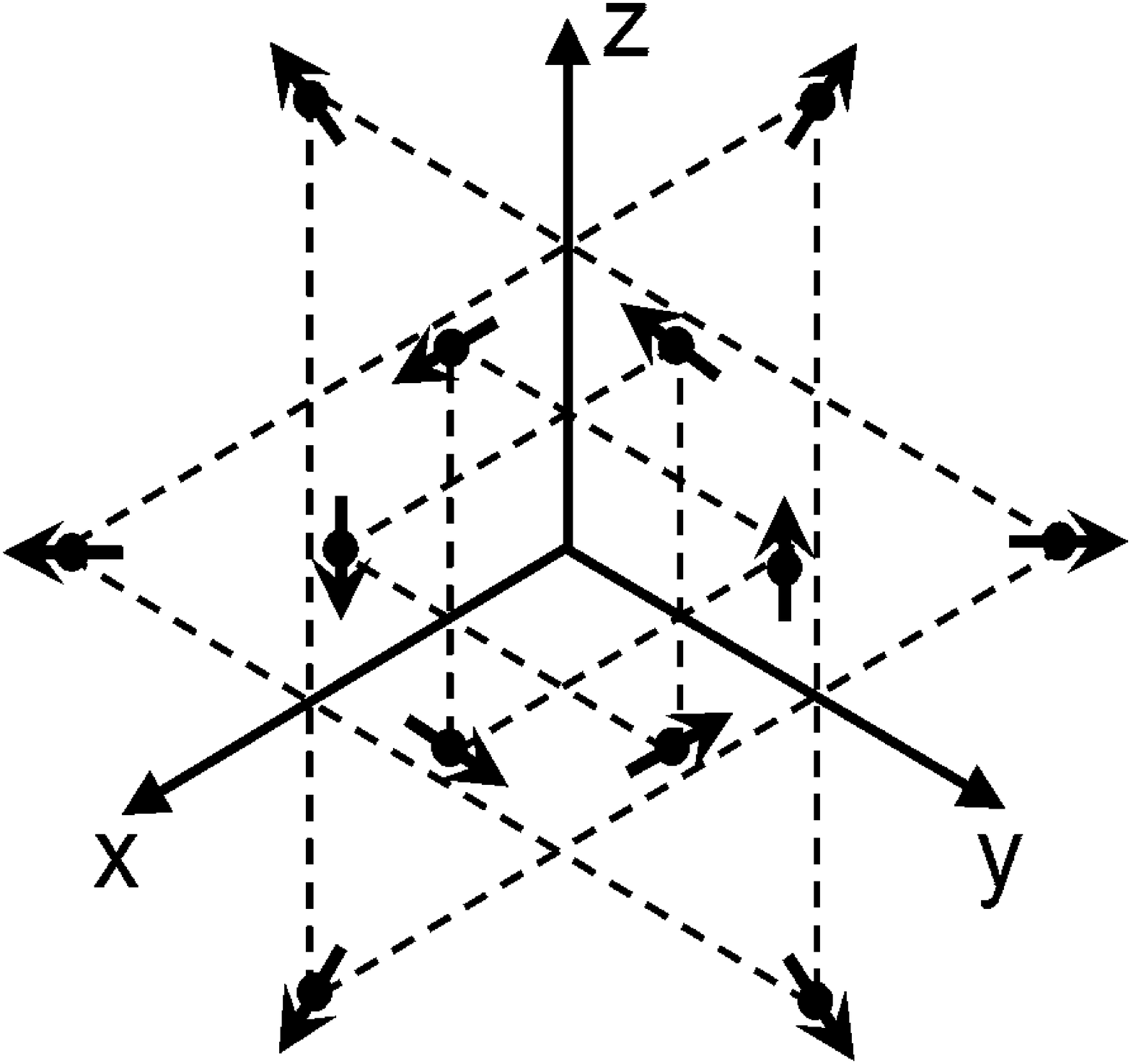, width=0.75\textwidth}

\vspace{0.1in}
Figure 5. Magnetic structure of noncollinear antiferromagnet
Mn$_3$Al$_2$Ge$_3$O$_{12}$.
\end{figure}

\begin{figure}
\hspace{0.5in}
\epsfig{file=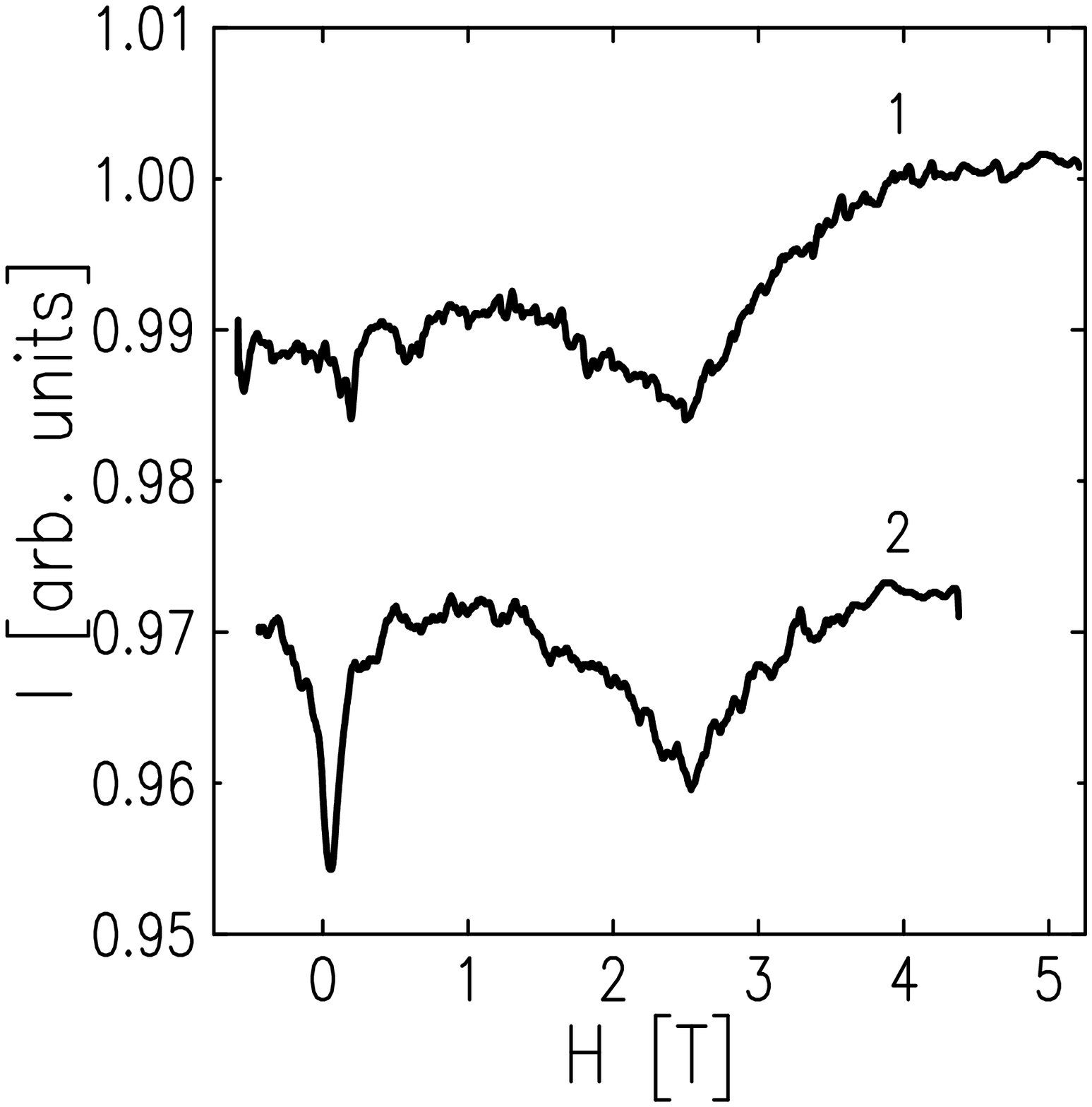, width=0.75\textwidth}

\vspace{0.1in}
Figure 6. Examples of NMR spectrum records at ${\bf h} \perp {\bf H}$ in
single crystal Mn$_3$Al$_2$Ge$_3$O$_{12}$ at $\approx 1.3$ K and {\bf H} $||$ [001]: 
(1) 604.4 MHz, coaxial resonance system and (2) 472.3 MHz, split-ring resonance system.
\end{figure}

\begin{equation}
d\Omega = \alpha P\left(U_f + I\int_{0}^{t}{U_fdt}+D\frac{dU_f}{dt} \right),
\end{equation}
where $P$, $I$, and $D$ are the feedback characteristics
dependent on temperature, $Q$, and other factors. Their values are selected manually; usually, $P >> I, D$.
Constant $\alpha$ is determined by the second derivative value of the amplitude - frequency characteristic of the
system on the top of the resonance peak.

To smooth the amplitude - frequency characteristic of the supplying high-frequency section, decoupling
attenuators At1 (10 dB) and At2 (3 dB) are placed at the input and output of the low-temperature part of the spectrometer. The output power of the RF generator is 0.1 W. The absorption in the resonance section is
detected by the second phase-sensitive voltmeter L2 (SR 830 lock-in amplifier) from variations of the second harmonic signal amplitude $U_{2f}$.

The spectrometer is intended to perform measurements in two modes: 
by scanning the magnetic field at a fixed generator frequency and by scanning 
the frequency at a fixed magnetic field. 
The need in this mode is caused by a weak dependence of the NMR
frequencies on the field when the dynamic frequency shift is absent. 
Unfortunately, due to a low operation speed of the digital AFC ($\sim$ 10 Hz), 
we failed to use in full this operation mode of the spectrometer.

By using the spectrometer, we studied the reorientation phase transition in a cubic crystal 
( $O^{10}_h$) of manganese garnet Mn$_3$Al$_2$Ge$_3$O$_{12}$, which is the 
noncollinear 12-sublattice antiferromagnet with Neel temperature $T_N$ = 6.8 K (see Fig. 5). 
According to neutron diffraction studies, Mn$^{2+}$ magnetic moments in magnetically ordered 
states are located in plane (111), being aligned with axes [211], [121], and [112] or
opposite to them. Thus, we obtain the noncollinear
triangular 12-sublattice antiferromagnet ordering \cite{13}. 
If the magnetic field is applied along direction [001], the spin plane rotates so that, 
in fields $H_c \sim 3$ T, it is oriented perpendicularly to the field.

Figure 6 shows records of NMR spectra ${\bf h} \perp {\bf H}$ in
the single crystal Mn$_3$Al$_2$Ge$_3$O$_{12}$, at $T \approx 1.3$ K, ${\bf H} ||$
[001], and (1) 604.4 MHz (the coaxial resonance system), and (2) 472.3 MHz 
(the split-ring resonance system). 
In all the scans, one can observe the minimum at $H_c = 2.5 \pm 0.3$ T, 
corresponding to the field of the phase transition. 
The appearance of the wide (in frequency) absorption line is obviously caused by
interaction with a low-frequency branch of the antiferromagnetic resonance \cite{14}. 
The value of the critical field $H_c$ of the reorientation transition at $T \approx 1.3$ K is
in agreement with magnetization measurement data in \cite{15}. 
In scan 2, the absorption in the vicinity of $H \sim 0$ is observed at frequencies below 
600 MHz and demonstrates a strong hysteresis behavior. 
It may indicate that antiferromagnetic domains inside the sample disappear. 
In our future experiments, we plan to obtain more detailed information about evolution
of the magnetic structure and low-frequency spin dynamics of Mn$_3$Al$_2$Ge$_3$O$_{12}$ 
in the magnetic field.

We are grateful to B.\,V.\,Mill for granting the manganese garnet single crystal and
to A.\,I.\,Kleev and V.\,I.\,Marchenko for the useful discussions.

\end{document}